\documentclass[12pt]{iopart}
\usepackage{graphicx}
\usepackage{amssymb}
\usepackage{psfig}

\begin{document}
\title[Collective decision making in cohesive flocks]
{Collective decision making in cohesive flocks}
\author{K. Bhattacharya$^{\,1}$ and
Tam\'as Vicsek$^{\,1,2}$
}
\address{$^{1}$ Department of Biological Physics, E\"otv\"os University, Budapest,
 H-1117 Hungary}
\address{$^{2}$ Statistical and Biological Physics Group, ELTE-HAS, 
P\'azm\'any P. Stny. 1A, Budapest, H-1117 Hungary}
\ead{tangkunal@gmail.com}
\begin{abstract}
Most of us must have been fascinated by the eye catching displays of collectively moving animals. Schools of fish can move in a rather orderly fashion and then change direction amazingly abruptly. There are a huge number of further examples both from the living and the non-living world for phenomena during which the many interacting, permanently moving units seem to arrive at a common behavioural pattern taking place in a short time. As a paradigm of this type of phenomena we consider the problem of how birds arrive at a decision resulting in their synchronized landing. We introduce a simple model to interpret this process. Collective motion prior to landing is modelled using a simple self-propelled particle (SPP) system with a new kind of boundary condition, while the tendency and the sudden propagation of the intention of landing is introduced through rules analogous to the random field Ising model in an external field. We show that our approach is capable of capturing the most relevant features of collective decision making  in a system of units with a variance of individual intentions and being under an increasing level of pressure to switch states. We find that – as a function of the few parameters of our model – the collective switching from the flying to the landing state is indeed much sharper than the distribution of the individual landing intentions. The transition is accompanied by a number of interesting features discussed in this report.
\end{abstract}
\pacs{87.15Zg, 87.19lo
}
\maketitle

\section{Introduction}
\label{intro}
In recent years, there has been a growing interest in the topic of consensus decision making in several fields including animal behaviour \cite{conradt-rev}, social sciences \cite{soc-science} and control theory \cite{control}. In this paper we use the paradigm of animal groups but we would like to stress already at this point that there is a considerable analogy between the decision-making processes in very different systems and we expect that our simple model should be applicable to a wider range of systems. Our object of study is the phenomenon during which the animals moving in groups, seemingly make unanimous decisions on the choice and time of performing activities even in the absence of global leaders. It is expected that because of the heterogeneity in the attributes like the age, sex and social status of the animals in a group or the differences in their perception of external stimuli there will be differences in the motivations of the members at the time of making choices. However, in spite of the differences, synchronization is seen to take place.  Examples include takeoff of swarm of honeybees from nest sites \cite{bee}, activity synchronization in sheep \cite{sheep}, collective movement in monkeys \cite{monkey}, group departures of domestic geese \cite{geese} and departure of ants from a feeding site \cite{ant}.

The study of collective opinion formation in physics has mainly focused on human societies \cite{castellano-rmp}. Parallels have been drawn between phase transitions in magnets and the opinion dynamics in populations. Alternatively, the subject of collective motion of animal groups has been extensively modelled \cite{spp,chate}. Here point-like particles, representing animals, move with constant speed, each tending to align with its immediate neighbours, for low noise levels, giving rise to a globally ordered state replicating the motion of flocks where all animals move in the same direction.

In this report we model the process of landing of bird flocks performing foraging flights as a typical example of collective decision making. We view this phenomena as a shift of the average opinion of the flock from that of continuing horizontal flight about some preferred altitude to that of descent towards the surface below.  We model the fact that birds land almost synchronously even in the presence of heterogeneity in motivations, along the lines of the random field Ising model (RFIM) \cite{galam,sethna}. Our approach is similar to that used in \cite{bouchaud} where the authors analyze (among other cases) the manner in which people in concert halls, initially joining the applause, stop clapping almost at the same time. In a very recent model \cite{daruka} for collective landing, birds are assumed to move under the action of different {\it social} forces. In addition, the internal state of each bird is characterized by a continuous variable called {\it landing intent} such that the internal state of each bird is directly coupled to the internal state of its neighbours. In contrast we allow motivation of individual birds to be influenced by only the observable variables of their neighbours, like position and velocity. In addition as a part of the model we introduce a method which enables us to work in open boundary conditions where the birds in flight move cohesively and do not spread out in the horizontal plane even in the presence of perturbations. With this model we investigate the level of synchronicity in collective landing across time and space and explore the nature of fluctuations close to the point in time when landing occurs.

\section{Model}
Below we describe the two aspects of the model, {\it viz.} collective decision making and collective motion in detail. The birds in our model are represented by particles characterized by the dynamical variables of position $(x,y,z)$ and velocity $(v^x,v^y,v^z)$ in three dimensional space. The only {\it natural} boundary that we allow in the system is the $z=0$ plane which we consider to be the landing surface. The $z$-coordinate of a flying bird gives its height above the ground. We assign to each particle $i$ a variable $a_i$ such that we refer to a bird in the flying state as active ($a_i=1$) and a bird in the landed state as inactive ($a_i=0$). This variable is updated at discrete times $t$ according to the condition:
\begin{equation}
\label{mi}
a_i(t)=\left\{ \begin{array}{ll}
1 & \textrm{if $z_i(t-\Delta t)>0$}\\
0 & \textrm{if $z_i(t-\Delta t)\le 0$ or $a_i(t-\Delta t)=0$,}
\end{array}\right.
\end{equation}        
where $\Delta t$ is the time increment. Once a bird $i$ has landed we keep its position unchanged till the end of the simulation. Also, from here on we use the words ``bird'' and ``particle'' synonymously.  

First we focus on the variables $\mathbf{x}^{\shortparallel}_i\equiv(x_i,y_i)$ and $\mathbf{v}^{\shortparallel}_i\equiv(v^x_i,v^y_i)$ governing the cruising motion of the birds parallel to the landing surface. This motion is assumed to be essentially decoupled from the motion in the vertical direction and is modelled in the spirit of \cite{spp}. A particle $i$ is assumed to move, with some uncertainty, in the average direction of motion of all neighboring particles $j$ whose separation $|\mathbf{x}^{\shortparallel}_j-\mathbf{x}^{\shortparallel}_i|$ from $i$ is less than a interaction radius $R$. From now on we refer to the set of particles in this neighbourhood as $\mathcal{N}_{i,R}$ with respect to a particle $i$ such that, the inactive particles and the particle $i$ are included as well \footnote{Neighbourhood could be defined in alternate ways as well. For example, Ballerini \etal \cite{ballerini-topological} found that for starlings, a better definition of the neighbourhood is based on rather the topological than the Euclidean distance}. The expression for updating velocity at a time $t$ is given by
\begin{equation}
\label{eom-xy}
\mathbf{v}^{\shortparallel}_i(t) = v \cdot
    \mathbf{N} \Bigg[
	\mathcal{M}(\xi^t_i) \cdot
	\mathbf{N} \Big(
	     \sum_{\stackrel{j\in\mathcal{N}_{i,R}}{a_j=1}}\mathbf{v}^{\shortparallel}_j(t-\Delta t)
	\Big)
	+ \mathbf{F}_{\mathrm{B}}\Big(\mathbf{x}_i^{\shortparallel}(t-\Delta t)\Big)
    \Bigg],
\end{equation}
where $v$ is the magnitude of the velocity,
$\mathcal{M}(\xi^t_i)$ is the rotation matrix in two dimensions
representing a random perturbation by an angle $\xi^t_i$ and $\mathbf{N}(\mathbf{u})$ denotes the unit vector $\mathbf{u}/|\mathbf{u}|$. Here $\xi^t_i$ is chosen with a uniform probability from the interval $[-\eta \pi,\eta \pi]$, where $\eta$ is the amplitude of the noise. We describe the force-like term $\mathbf{F}_\mathrm{B}$ below. We follow the standard version  \cite{huepe-aldana-1} of the original SPP model \cite{spp} for the updating of positions:
\begin{equation}
\mathbf{x}^{\shortparallel}_i(t)=\mathbf{x}^{\shortparallel}_i(t-\Delta t)+\mathbf{v}^{\shortparallel}_i(t)\Delta t.
\label{eq_snm_pos}
\end{equation}

\begin{figure}
\begin{center}
\includegraphics[width=9.0cm]{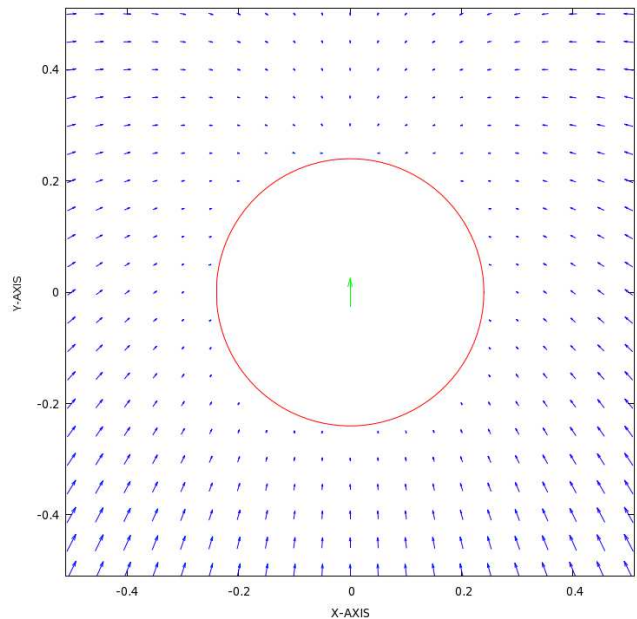}
\end{center}
\caption{A typical snapshot of the vector field $\mathbf{F}_{\mathrm{B}}(\mathbf{x}^{\shortparallel})$ on the $xy$-plane, defined through (\ref{bound-1}) for the value of $\beta=0.75$. The direction of motion of the {\it CoM} is assumed to be along the Y-axis, indicated by the arrow at the centre of the circle. The length of a vector is directly proportional to the distance from the boundary.
}
\label{field}
\end{figure}

We mention here that it is a well-known problem with existing models of collective motion to maintain the cohesiveness of a moving flock in the presence of a finite amount of noise $\eta$ and with finite ranged (as $R$) interaction between the members. A flock tends to break up into smaller sub-flocks which eventually move in independent directions. A periodic boundary condition (PBC) is used to prevent the flock from spreading out perpendicular to the direction of motion. However, in the problem of landing we find that the imposition of a PBC would be unrealistic in the sense that there will be a possibility of a particle that has become inactive and hence ceased to move, to continually exchange information with the rest of the moving flock. Therefore, we introduce a new kind of ``comoving boundary'' condition where we imagine the motion of the particles on the $xy$-plane to be governed by a circular bounding region with radius $R_\mathrm{B}$ and the centre lying on the centre of mass ({\it CoM}) of all active particles (mass of each assumed to be unity). This boundary therefore translates with the flock as the {\it CoM} is put into motion due to collective motion of the particles. When a particle which is inside the region tries to leave it, the particle is subjected to an attraction proportional to the distance from the boundary. The direction of this attraction is along the resultant of the direction towards the location of the {\it CoM} and the direction of motion of the {\it CoM} itself. For a particle located at $\mathbf{x}^{\shortparallel}$ at an instant of time, the attraction $\mathbf{F}_\mathrm{B}(\mathbf{x}^{\shortparallel})$ is given by:
\begin{equation}
\label{bound-1}
\mathbf{F}_\mathrm{B}(\mathbf{x}^{\shortparallel}) = \left\{
    \begin{array}{ll}
      D_\mathrm{B}  
    	    
       \Big[
       |\mathbf{x}^{\shortparallel}-\mathbf{\overline{x}}^{\shortparallel}| - R_\mathrm{B} 
     \Big]\mathbf{C}\big(\mathbf{x^{\shortparallel}}\big)
        & \mbox{ if $|\mathbf{x}^{\shortparallel}-\mathbf{\overline{x}}^{\shortparallel}| > R_\mathrm{B}$ }\\
      0 & \mbox{ otherwise, }
    \end{array}
\right.
\end{equation}
with
\begin{equation}
\label{bound-2}
\mathbf{C}(\mathbf{x}^{\shortparallel})=-\mathbf{N}\big(\mathbf{x}^{\shortparallel}-\mathbf{\overline{x}}^{\shortparallel}\big)+\beta\mathbf{N}\big(\overline{\mathbf{v}}^{\shortparallel}\big),
\end{equation}
where $D_\mathrm{B}$ is the strength of the attraction, $\mathbf{\overline{x}}^{\shortparallel}$ and $\mathbf{\overline{v}}^{\shortparallel}$ are the position and the velocity of the {\it CoM}, respectively, parallel to the $xy$-plane at that instant and $\beta$ is a parameter. The {\it CoM} and its velocity are calculated with respect to the active particles. In the figure \ref{field} we plot $\mathbf{F}_\mathrm{B}$ when the motion of the {\it CoM} is in a direction parallel to the Y-axis. Similarly, we assume that the boundary adjusts itself to keep average separation between active particles constant. Thus, in general $R_\mathrm{B}$ is a function of time, given by, 
\begin{equation}
\label{rb}
R_{\mathrm{B}}(t)=\sqrt{\frac{NA(t)}{\pi\rho}},
\end{equation}
where $A(t)=\sum_i a_i(t)/N$ is the fraction of the active particles at time $t$ and $\rho$ is the number of particles, initially active, per unit area of the $xy$-plane.

Now we explain how the collective decision making process takes place in the flock and how it is related to the cohesion and eventual landing along the vertical direction. We characterize the decision of an individual bird as a binary choice problem \cite{bouchaud}. To each bird $i$ we assign an internal state variable $s_i$ such that when $s_i=1$ the bird continues to cruise above ground and when $s_i=-1$ the bird decides to land. For the cruising state we assume that a bird likes to fly about an altitude $z_0$ on average without deviating too far above or below. However, we expect the elevation of a bird to fluctuate with time pertaining to factors like the noise in the medium and collision avoidance. Thus when $s_i=1$ we consider the vertical motion of the particle to be essentially a random walk bounded by imaginary walls at $z=z_0-\frac{\Delta z_0}{2}$ and $z=z_0+\frac{\Delta z_0}{2}$ where $\Delta z_0$ is the thickness of the flock. In case the bird decides to land the vertical motion is directed towards the landing surface at $z=0$. Formally we define,
\begin{equation}
\label{spin}
v^z_i(t)=\left\{ \begin{array}{ll}
\phantom{x}\epsilon^t_i\cdot v & \textrm{if $s_i(t)=1$ and $|\delta(t-\Delta t)|\leq \Delta z_0/2$}\\
-v\cdot \textrm{sign}\big[\delta(t-\Delta t)\big] & \textrm{if $s_i(t)=1$ and $|\delta(t-\Delta t)|>\Delta z_0/2 $}\\
-v &  \textrm{if $s_i(t)=-1$,}
\end{array}\right. 
\end{equation}   
where $\epsilon^t_i$ takes values $\pm 1$ with equal probability and $\delta(t-\Delta t)=z(t-\Delta t)-z_0$, is the amount of deviation from the height $z_0$ at time $t-\Delta t$. 

To model the motivational differences in birds we assign to each bird $i$, an inherent switching time $t_i$ \cite{conradt-nature} such that if the bird begins an isolated flight at time $t=0$, it would decide to land at time $t=t_i$. We assume that the value of $t_i$ for the bird $i$ will in general depend on its energy reserves \cite{sirot} and thus in general will be different for different birds. We choose $t_i$'s from a Gaussian distribution with a given standard deviation $\sigma_0$. The value of $\sigma_0$ allows us to control the level of heterogeneity in a flock. Although we assign {\it a priori} values of $t_i$ to birds, due to the randomness in the choice of these switching times, they can be thought to be additionally include the effect of the environment such as quality of foraging patches over which the flocks fly.

For a particle in the flock, it will have a finite number of particles in its neighbourhood and therefore will be influenced by their actions while taking a decision. Hence the internal state of a particle will be a function of two competing factors, (i) its inherent switching time and (ii) the internal state of the neighbours, which we assume to be reflected by the nature of their motion. In an approach similar to that of the RFIM at zero temperature \cite{bouchaud} we write down:
\begin{equation}
\label{rfim}
s_i(t)=\textrm{sign}\Big [t_i-t+J\bar{S}_{\mathcal{N}_{i,R}}(t-\Delta t)\Big ]
\end{equation}    
where $J$ is the propensity of a particle to follow the decision of its neighbours and $\bar{S}_{\mathcal{N}_{i,R}}$ is the {\it local mean field} through which the bird $i$ senses the average decision in its neighbourhood $\mathcal{N}_{i,R}$. We expect that the inclination of a bird $i$ to land will be enhanced by a resultant motion of birds in $\mathcal{N}_{i,R}$ towards the landing surface, and also by the number of birds that have already landed. We define $\bar{S}_{\mathcal{N}_{i,R}}$ at a time $t$ as:
\begin{equation}
\label{mean-field}
\bar{S}_{\mathcal{N}_{i,R}}(t)=1+2A_{\mathcal{N}_{i,R}}(t)\frac{\big\langle v^z(t)\big\rangle_{\mathcal{N}_{i,R}}}{v}-2\Big(1-A_{\mathcal{N}_{i,R}}(t)\Big),
\end{equation}
where $\big\langle v^z(t)\big\rangle_{\mathcal{N}_{i,R}}$ and $A_{\mathcal{N}_{i,R}}(t)$ are the average value of $z$-component of velocity of active particles and fraction of active particles, respectively, in the neighbourhood $\mathcal{N}_{i,R}$ at time $t$. At times $t<<\langle t_{i}\rangle-\sigma_0$, the birds on average have no inclination to land and therefore we expect them to maintain a height around $z=z_0$ without any resultant vertical motion towards the landing surface. In this regime we expect for any particle $i$, $\big\langle v^z\big\rangle_{\mathcal{N}_{i,R}}\sim 0$ and also, $1-A_{\mathcal{N}_{i,R}}=0$, implying that  $\bar{S}_{\mathcal{N}_{i,R}}\sim 1$. On the other hand a resultant vertical motion in negative $z$-direction would imply $\big\langle v^z\big\rangle_{\mathcal{N}_{i,R}}\to -v$. Thus in a regime when majority move towards the landing surface at $z=0$ and eventually reach inactive states, we expect $\bar{S}_{\mathcal{N}_{i,R}}\to -1$. We now move on to characterize the resulting collective landing as a function of different states of collective motion. {\it We would like to emphasize at this point that although the detailed description of our model involves specifying nine expressions, the essential concept and the ingredients of the model are rather simple (SPP + RFIM) and the number of independent parameters is low ($\eta$, $R$, $\sigma$, $J$)}.

\section{Results}   
We note that in our model, the shape of a coherently moving flock, far from the landing regime, is roughly cylindrical with a radius of cross section $R_\mathrm{B}(0)=\sqrt{N/\rho}$ and thickness $\Delta z_0$. We argue that from the point of view of our model for large flocks $\Delta z_0$ will have in general weak or no dependence on the number of birds.  We choose the thickness of the flock $\Delta z_0$ as the unit to measure distances ($\Delta z_0=1.0$). We choose the projected density $\rho=2.0$, the interaction radius $R=2.0$ and the preferred altitude of flying $z_0=10.0$. In our simulations we take $N=1024$, unless mentioned otherwise, the order of which is typical for moderately large flocks \cite{large-flock}. Also the ratio $\Delta z_0/z_0$ would be typical in these flocks which have thickness of order $10$ m and fly at altitudes of around $100$ m. At the time $t=0$ in (\ref{rfim}) we expect the flock to be already in the steady state of collective motion. So before we switch on time as in (\ref{rfim}) we evolve a flock to its steady state of motion according to equations (\ref{eom-xy}), (\ref{eq_snm_pos}) and (\ref{spin}) where each particle $i$ has $s_i=1$. It is known \cite{spp,chate} that the nature of the collective motion depends on the amplitude of the noise. We find that for sufficiently low values of $\eta$, all the particles align their directions and the flock as a whole translates in some spontaneously chosen direction parallel to the $xy$-plane. As the value of $\eta$ is increased, the tendency of alignment is gradually lost leading to weak or no translation. However, the actual nature of this transition is not known precisely enough and its detailed investigation is outside of the scope of the
present paper.

\subsection{Time scales}
We make an observation that there are two independent time scales in the problem - the time scale of motion and the time scale over which the inherent switching times are distributed. The former is given by $T_v=\Delta z_0/v$ which is the time required to move one unit of length (or half interaction radius $R=2$). In presence of finite amount of noise $\eta$, this time scale would also govern the mixing between particles. The width of the distribution of $t_i$'s ($\approx 2\sigma_0$) gives a measure of the time over which the energy reserves of the majority in the flock are exhausted. Therefore, when $\sigma_0\lesssim T_v$, there is hardly any motion during the collective decision making. When $\sigma_0\gg T_v$ the flock actually translates (at low values of $\eta$) over a long distance during the process. We call the ratio of these two time scales as $\lambda_0$, where $\lambda_0=\sigma_0/T_v=\sigma_0 v/\Delta z_0$, and we investigate the nature of landing at its different values.

We set the time scale for choosing values  of $v$ and $\sigma_0$ as the time increment $\Delta t=1.0$. The value of magnitude of velocity $v$ is kept around and below $0.1$ so that change in the set of neighbours $\mathcal{N}_{i,R}$ of any bird $i$ is gradual. We choose the value of the parameters for the comoving boundary described by equations (\ref{bound-1}) and (\ref{bound-2}) to be $D_\mathrm{B}\sim 10^{-3}$ and $\beta = 0.75$. We also observe that parameter $\beta$ controls the degree of asymmetry of the vector field $\mathbf{C}$ (\ref{bound-2}) outside the comoving boundary. For $\beta\rightarrow 0$ the system enters into a rotating phase \cite{erdmann}, and for $\beta\gg 1$ the flock has a tendency to accumulate at the boundary. 

\subsection{The landing process}
\begin{figure}
\begin{center}
\includegraphics[width=15.0cm]{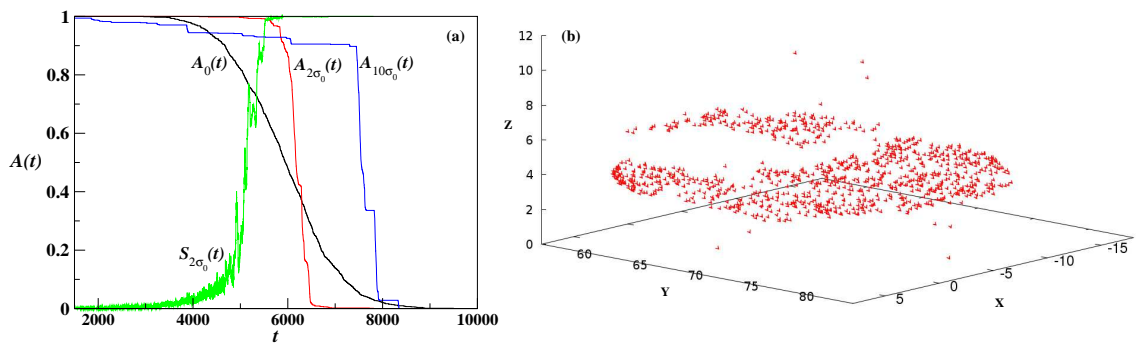}
\end{center}
\caption{(a) The quantity $A(t)$ denotes the fraction of birds not landed yet as the time $t$ progresses. The curves $A_0(t)$, $A_{2\sigma_0}(t)$ and $A_{10\sigma_0}(t)$ correspond to values of coupling $J=0, 2\sigma_0$ and $10\sigma_0$, respectively. Also, $S_{2\sigma_0}(t)$ shows the fraction of birds in state $s_i=-1$ at time $t$ for $J=2\sigma_0$. (b) A snapshot of a collective landing of a bird flock with $N=1024, \rho=2.0, \eta=0.2, R=2.0$ and $v=0.01$ with $\sigma_0=1000$ and $J=2\sigma_0$. The arrowheads point in the direction of the velocity of the birds.
}
\label{trans-land}
\end{figure}

The role played by the coupling $J$ in (\ref{rfim}) is well known \cite{bouchaud,sethna} and has also been the matter of investigation. However, in our model we fix $J$ to an optimal value for which we find the desired behaviour, as explained below. In the figure \ref{trans-land}(a) we compare the process of landing across time for different values of the coupling. For a flock with inherent switching times $\{t_i\}$ chosen with a fixed $\sigma_0$ we monitor the fraction of active particles $A_J(t)$ with time $t$ for different values of $J$. We observe that when $J=0$ the particles become inactive independent of each other at their own switching times as seen from the curve $A_0(t)$ which falls smoothly with time. On choosing $J=10\sigma_0$ the behaviour is quite opposite. A very sharp fall occurs in the curve $A_{10\sigma_0}(t)$ implying that there is indeed a collective decision making taking place when a large number of particles reach inactive states within a short period of time. However, comparing the two curves $A_0(t)$ and $A_{10\sigma_0}(t)$ it appears that in the latter case, the collective landing occurs at point in time which is much larger than (by an amount of order $\sigma_0$) than the switching times of most of the birds. We find that a much lower value of $J=2\sigma_0$ is relevant biologically. The curve $A_{2\sigma_0}(t)$ sharply falls indicating collective behaviour and the fall occurs near at a time when about half of the birds have just crossed their inherent switching times. 

In the figure \ref{trans-land}(a) we also plot $S_{2\sigma_0}(t)$ which is the fraction of birds which have a momentary intention to land at time $t$ {\it i.e.}, the fraction of particles with $s_i=-1$. Unlike $A_{2\sigma_0}(t)$, the quantity $S_{2\sigma_0}(t)$ is seen to fluctuate before becoming equal to the maximum value unity. This shows that before finally deciding to land, birds may intermittently change their decision. This is caused by the fluctuations in the vertical motion (\ref{spin}) of neighbours as well as the changing neighbourhood due to finite value of $\eta$. However, these fluctuations become only important near the landing regime. We explore an aspect of these fluctuations later. In the figure \ref{trans-land}(b) we plot the velocity field of a typical flock when the majority has decided to land.
\begin{figure}
\begin{center}
\includegraphics[width=15.0cm]{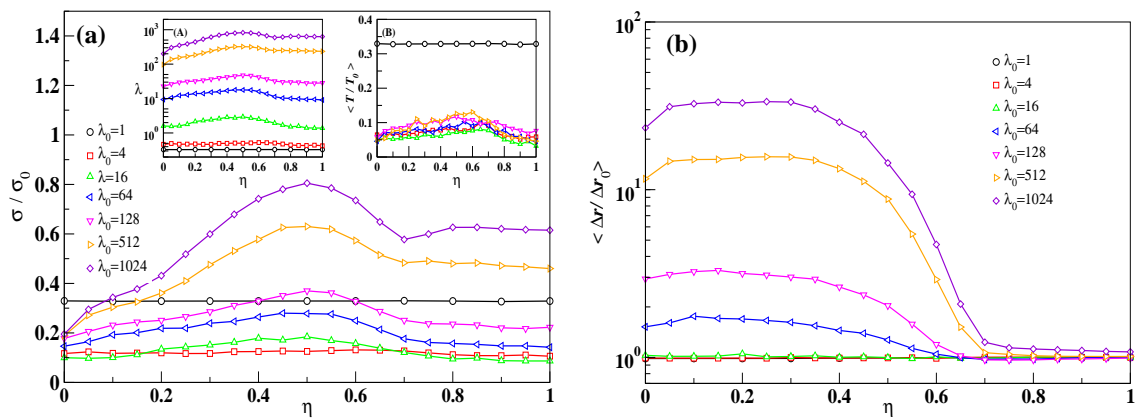}
\end{center}
\caption{
(a) Plot of the ratio of the standard deviation ($\sigma$) of actual landing times to the standard deviation ($\sigma_0$) of inherent switching times as a function of noise $\eta$ for different values of $\lambda_0$, {\it i.e.}, $\sigma_0$ scaled by the scale of motion $T_v$, as indicated by different symbols. Inset (A) shows variation of $\lambda$ {\it i.e.}, $\sigma$ scaled by $T_v$, with noise. Inset (B) shows the variation of $\langle T/T_0\rangle$ which ignores the first and last $20\%$ of the landings. (b) The variation of the spread of the landed flock over the $xy$-plane, relative to the spread at time $t=0$, as a function of noise.       
}
\label{t2080-r}
\end{figure}

\subsection{Role of parameters}
Below we quantify the extent to which the presence of the coupling affects the collectiveness in landing as we vary the scaled value of the  heterogeneity ($\lambda_0$). We define a measure of the sharpness of the decrease in the fraction of active particles $A(t)$ (figure \ref{trans-land}). During the process of landing, we record the times at which individual particles become inactive. Then we calculate the standard deviation $\sigma$ of the distribution of these landing times. Finally, we divide $\sigma$ by $\sigma_0$ to get the normalized measure. The dimensionless quantity $\sigma/\sigma_0$ averaged over different realizations of ${t_i}$'s, provides a measure of the extent to which the landing of the flock is coherent in time such that $\sigma=\sigma_0$ implies the complete absence of collective decision making. In the figure \ref{t2080-r}(a) we plot $\sigma/\sigma_0$ versus $\eta$ for different values of $\lambda_0$. We make the following observations. For $\lambda_0\simeq 1$ the flock reaches consensus almost in the absence of any motion. As a result there is no variation of $\sigma/\sigma_0$ with the magnitude of noise. In this regime the value of $\sigma$ is governed by the spread of the flock in the steady state along the vertical direction {\it i.e.}, $\Delta z_0$. Therefore we expect that $\sigma\sim \Delta z_0/v$. In the inset (A) of the figure \ref{t2080-r}(a) we plot the ratio $\lambda=\sigma/(\Delta z_0/v)$ against $\eta$. It is worth mentioning that the quantity $\lambda$ provides an alternate normalization of $\sigma$ with respect to the other independent time scale $T_v$($=\Delta z_0/v$). We find that for $\lambda_0$ close to unity, $\lambda<1$ asserting that the landing is coherent and the separation between landing times is due to the width $\Delta z_0$. For larger values of $\lambda_0$ the value of $\sigma/\sigma_0$ also becomes large signifying gradual loss of coherence. We also observe that in general $\sigma/\sigma_0$ has a maximum for intermediate values of the noise. 

We find that if we choose to ignore the particles which become inactive either too early or too late, then the level of temporal coherence appears better. For a particular landing we measure the times $t_{20}$ and $t_{80}$ at which $20\%$ and $80\%$ of the particles, respectively, in the flock has just become inactive. Thus we define, $T=t_{80}-t_{20}$ as the width of the time window in which a majority {\it i.e.}, $60\%$ of the particles become inactive. We normalize $T$ by the corresponding time window ($T_0$) for a flock with the coupling $J=0$. In the inset (B) of the figure \ref{t2080-r}(a) we plot $\langle T/T_0\rangle$ versus $\eta$ for different values of $\lambda_0$. The average behaviour reveals a level of coherence such that two birds originally having their inherent switching times separated by $T_0$ land within a period of around $7\%$ of $T_0$.

We investigate the degree of collectiveness as a function of the distance over which a flock lands in figure \ref{t2080-r}(b). We calculate the following quantity:
\begin{equation}
\Delta r=\sqrt{\frac{\sum_i\big(\mathbf{x}^{\shortparallel}_i\big)^2}{N}-\bigg(\frac{\sum_i\mathbf{x}^{\shortparallel}_i}{N}\bigg)^2}.
\end{equation}           
This quantifies the spread of the flock across the landing surface. We normalize the value of this spread ($\Delta r$) when all particles become inactive by the corresponding value ($\Delta r_0$) calculated with the positions of the active particles at time $t=0$. In the figure \ref{t2080-r}(b) we plot $\langle\Delta r/\Delta r_0 \rangle$ as function of noise varying values of $\lambda_0$. For low and moderate values of $\lambda_0$, the value of $\langle\Delta r/\Delta r_0 \rangle$ is of the order of unity. The landing occurs over a region whose dimensions are comparable to dimensions of the flock.

We observe that in the regime of large $\lambda_0$, the value of noise plays a major role in deciding the overall nature of landing. For very low values of $\eta$ the flock translates over large distances. Due to this translation, the particles reaching inactive states early are unable to influence the decision of the active particles in the flock as the {\it CoM} moves away from these particles. In this case the landing occurs in few large clusters separated in space and time. In case of higher values of the noise there is weak or no translation of the {\it CoM}. However, the scale of the difference in inherent switching times {\it i.e.}, $\sigma_0$ being very slow as compared to the scale of motion, it provides the scope for a particle with a low value of $t_i$ to diffuse to the boundary. As a result, temporarily the influence of the neighbourhood is weakened. This allows the particle to switch its state and become inactive. Initially, the excursions of the particles to the comoving boundary becomes the reason behind independent landings. The shrinking of the boundary (\ref{rb}) effectively screens (depending upon the interaction radius $R$) the active particles from the inactive particles. For moderate values of $\eta$ the path of the {\it CoM} becomes tortuous. However, the effect of segregation of inactive particles from the active particles still persists. The landing occurs in large number of smaller clusters.

We further characterize the above behaviour by the following definition. After the landing we calculate the {\it CoM} of the particles positioned on the $xy$-plane. We define the measure of spatial coherence $\Psi$ as the fraction of particles that lie within a distance $R_{\mathrm{B}}(0)$ from the {\it CoM}. Formally,
\begin{equation}
\Psi=\frac{1}{N}\sum_{i=1}^{N}H\Big(R_{\mathrm{B}}(0)-|\mathbf{x}^{\shortparallel}_i-\mathbf{\overline{x}}^{\shortparallel}|\Big),
\end{equation} 
where, $\mathbf{x}^{\shortparallel}_i$ is the position of the $i$-th inactive particle, $\mathbf{\overline{x}}^{\shortparallel}$ is the position of the {\it CoM} and $H(...)$ is the Heaviside step function. We expect from the above observations that $\Psi$ is close to unity for a spatially coherent landing and zero in the opposite case. 

\begin{figure}
\begin{center}
\includegraphics[width=15.0cm]{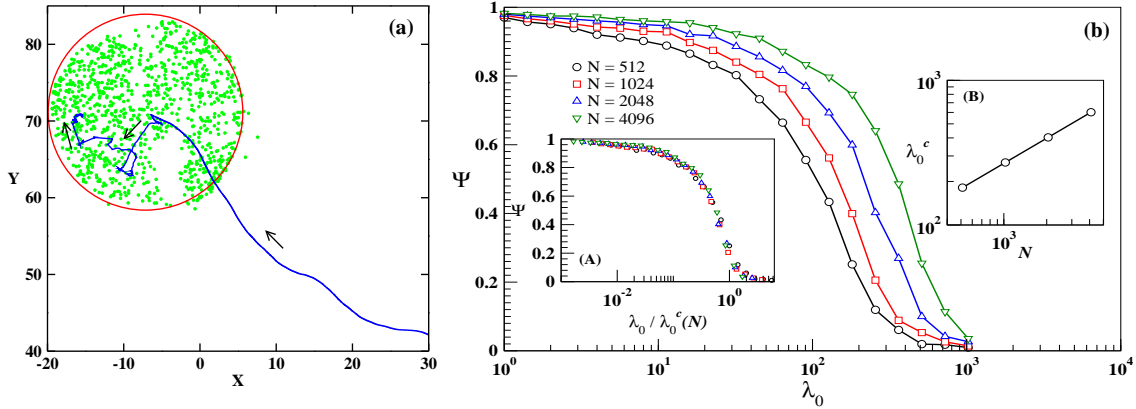}
\end{center}
\caption{(a) The snapshot after the flock, described in the figure \ref{trans-land}(b), has landed. The line shows the trajectory of the centre of mass of the birds in flight. The arrows indicate the average direction of motion of the centre of mass at different points in time. The circle drawn is centred on the centre of mass of all landed birds and is drawn with the radius $R_\mathrm{B}(0)$ of the comoving boundary. (b) The measure of spatial coherence $\Psi$ as a function of $\lambda_0$ for the value of $\eta=0.5$. Different symbols indicate the system sizes $N=512$, $1024$, $2048$, and $4096$. Inset (A) shows the plots with $\lambda_0$ rescaled to $\lambda_0/\lambda_0^c(N)$. Inset (B) shows the variation of $\lambda_0^c(N)$ with system size $N$.}
\label{com-psi}
\end{figure}

\begin{figure}
\begin{center}
\includegraphics[width=15.0cm]{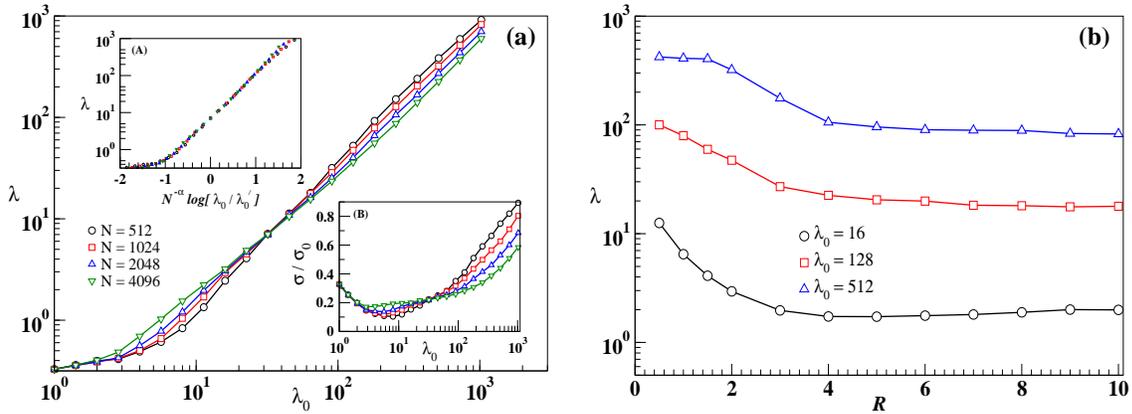}
\end{center}
\caption{(a) The scaled values ($\lambda$) of the standard deviation of the actual landing times as function of the scaled values ($\lambda_0$) of the standard deviation of the assigned switching times for different system sizes ($N=512$, $1024$, $2048$ and $4096$) indicated by different symbols. Inset (A) shows the collapse of the previous plots as result of transformation of the horizontal axis from $\lambda_0$ to $N^{-\alpha}\textrm{log}(\lambda_0/\lambda_0')$ with $\alpha=0.1$ and $\lambda_0'=32.0$. Inset (B) shows the variation of the ratio $\sigma/\sigma_0$ with $\lambda_0$. (b) The variation of $\lambda$ as a function of the interaction radius $R$ for different values of $\lambda_0=16$, $128$ and $512$ as indicated by different symbols, in flocks of size $N=1024$. The value of $\eta$ in both (a) and (b) is $0.5$.}

\label{t-r}
\end{figure} 

In the figure \ref{com-psi}(a) we plot the positions of the birds on the $xy$-plane after all the birds have landed. The line shows the trajectory of the {\it CoM} and the circle is drawn with the radius $R_\mathrm{B}(0)$ about the centre of mass calculated with the final positions of the landed birds. Majority of the birds lying within the circle shows that in such a case spatial coherence is retained. In the figure \ref{com-psi}(b) we plot $\Psi$ as function of $\lambda_0$ for the value of $\eta=0.5$. The different plots corresponds to different system sizes. The decrease in $\Psi$ with increase in $\lambda_0$ is consistent with all the previous observations. In the inset (A) of the figure \ref{com-psi}(b) we obtain a collapse of the data after rescaling $\lambda_0$ by $\lambda_0^c(N)$. The variation of $\lambda_0^c$ with the system size $N$, as shown in the inset (B), is found to be $\lambda_0^c\sim N^\gamma$ where $\gamma\simeq 0.6$ (we note that the dimension of the flock varies as $N^{0.5}$). This relation suggests that with the increase in the system size there is an increase in the value of $\lambda_0$ below which spatial coherence is expected. We investigate the dependence on system size of the temporal behaviour in the next paragraph.

In the figure \ref{t-r}(a) we plot $\lambda$ versus $\lambda_0$ for four different system sizes for value of noise $\eta=0.5$. The figure reveals how the spread in actual landing times increases with the value of $\lambda_0$. A data collapse in the inset (A) reveals a scaling relation of the form:
\begin{equation}
\lambda\sim\Big[\frac{\lambda_0}{\lambda_0'}\Big]^{\frac{1}{N^\alpha}},
\label{lambda-scale}
\end{equation}   
with $\alpha=0.1$ and $\lambda_0'=32.0$. In general, we expect $\lambda_0'$ to be function of the parameters like $\rho$ and $R$. The equation (\ref{lambda-scale}) suggests that for a fixed value of $\lambda_0$, with the increase in the system size there is increase in temporal coherence. We also plot the corresponding values of the ratio $\sigma/\sigma_0$ in the inset (B). The plot shows how this ratio approaches the maximum value unity with the increase in the value of $\lambda_0$. The reason for the rise in $\sigma/\sigma_0$ for very low values of $\lambda_0$ is already stated with respect to the figure \ref{t2080-r}(a). In all the above investigations we have assumed that the interaction radius controlling the average number of neighbours of any particle is $R=2.0$. We now address the issue of having larger (or smaller) interaction radii. In the figure \ref{t-r}(b) we plot the variation of $\lambda$ with the value of $R$ for different values of $\lambda_0$ with value of noise set at $\eta=0.5$. We find significant decrease in the spread $\lambda$ for large values of $R$. This suggests that global information can lead to a more coherent decision making process.

\begin{figure}
\begin{center}
\includegraphics[width=14.0cm]{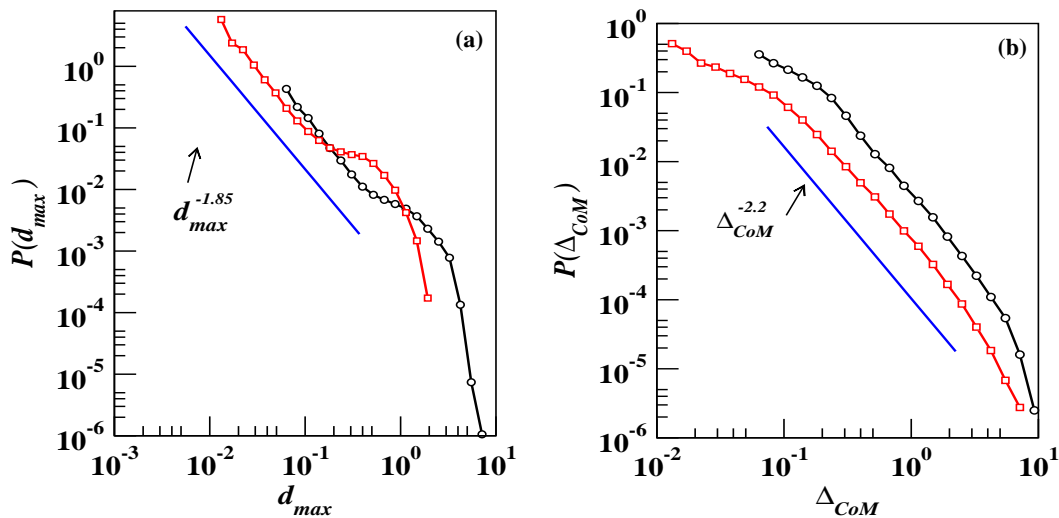}
\end{center}
\caption{(a) The probability density function $P(d_{max})$ of the maximum depth $d_{max}$ of excursions below the flock by birds, before returning to the flock as a result of temporarily reverting from the decision to land. The parameter sets indicated by symbols are $v=0.05$, $\sigma_0=200$ (circles) and $v=0.01$, $\sigma_0=1000$ (squares). The solid line has a slope $-1.85$ and is a guide to the eye. (b) The probability density function $P(\Delta_{{\it CoM}})$ of the distance travelled by the centre of mass of birds in flight, $\Delta_{{\it CoM}}$ between consecutive time steps. The parameter sets indicated by symbols are same as in the case of (a). The measured slopes of the power-law fits to the curves are around $-2.2$. In both (a) and (b) the value of $\lambda_0=\sigma_0 v/\Delta z_0=10$, $\eta=0.2$ and the magnitude of the velocity $v$ controls the lower cut-off of the power-law decays.}
\label{aval-dcm}
\end{figure}

\subsection{Fluctuations}
Lastly, we focus on the nature of fluctuations during the landing process. As a result of the decision reversals in birds under the influence from the neighbourhood (as reflected by $S_{2\sigma_0}(t)$ in the figure \ref{trans-land}(a)) we find that birds originally moving towards the landing surface sometimes return back to the cruising altitude. We characterize these fluctuations by the maximum depth $d_{max}$ to which a bird makes an excursion before returning to an altitude greater than $z_0-\Delta z_0/2$. In the figure \ref{aval-dcm}(a) we plot the probability density $P(d_{max})$. We find that the distribution of $d_{max}$ is a power-law decay bounded by a  minimum and maximum governed by the magnitude of velocity $v$ and the maximum possible depth $z_0$, respectively. We also study the jumps that occur in position of the {\it CoM} (on the $xy$ plane) of flying birds once the landing process starts. As is seen from the figure \ref{com-psi}(a) the trajectory of the {\it CoM} becomes irregular near the end. The {\it CoM} actually jumps from a cluster which ceases to move to a cluster which continues moving. We construct the probability density $P(\Delta_{{\it CoM}})$ for the separation $\Delta_{{\it CoM}}$ between the centres of mass at two consecutive time steps. We find that $P(\Delta_{{\it CoM}})$ has a fat tail during the landing process. In the figure \ref{aval-dcm}(b) we plot the tail of $P(\Delta_{{\it CoM}})$ after binning. The plots reveals power law decays with close by exponents. The power law has a cut-off which governed by $R_\mathrm{B}(0)$. 
 
\section{Conclusion}
In this paper we have investigated the interplay between the dynamics of collective motion and collective decision making. In particular we have presented a model for landing of bird flocks. We have identified the relevant scales in such a problem and we have worked in the coupling regime which is expected to be biologically relevant. It appears that higher values of the coupling delays the process of the global opinion shift. We have characterized the degree of the resulting collective behaviour by different measures. We found that --- as a function of the few parameters of our model --- the collective switching from the flying to the landing state is indeed much sharper than the distribution of the individual (inherent) landing intentions. Among the several interesting features of the
model we proposed, we showed that if the flock is subject to larger fluctuations, the spatial spreading of the fully landed flock becomes smaller due to a better mixing of the information about the momentary landing intentions of the birds. We also find that there is a characteristic value of heterogeneity which scales with the size of the flock. Below this value of heterogeneity, flocks remain spatially coherent in the process of landing and above this value, the coherence is lost. In addition we have formulated a boundary condition which is pertinent to models of flocking. We believe that the approach we proposed is also relevant for many other systems where abrupt behavioural changes are observed. Possible applications include various kinds of animal or even robotic groups in which an almost instantaneous global switching to a new desired state takes place.


\end{document}